\documentclass[conference,a4paper]{IEEEtran}

\usepackage[utf8]{inputenc} 
\usepackage[T1]{fontenc}
\usepackage{url}
\usepackage{ifthen}
\usepackage{cite}
\usepackage[cmex10]{amsmath} 
	\usepackage{amssymb, graphicx, url}
\usepackage{acronym}
\usepackage{times,bbm}
\usepackage{algorithm,algorithmic,bbm}
\usepackage{color}
\usepackage{wrapfig}
\usepackage{mwe}
\usepackage[final]{pdfpages}
\usepackage{subcaption}

\newfont{\bbb}{msbm10 scaled 500}

\newfont{\bb}{msbm10 scaled 1100}

\newcommand{\EE}{\mathbb{E}}
\newcommand{\PP}{\mathbb{P}}
\newcommand{\indic}{\mathbbm{1} }

\newcommand{\var}{{\hbox{var}}}

\newcommand{\bp}{\noindent{\textbf{Proof.}}\ }
\newcommand{\ep}{\hfill $\Box$}

\newcommand{\el}{\end{flushleft}}
\newcommand{\bl}{\begin{flushleft}}

\newcommand{\separator}{
	\begin{center}
		\rule{\columnwidth}{0.3mm}
	\end{center}
}

\newcommand{\limK}{ {\underset{K \to \infty}{\lim}} }
\newcommand{\PtoK}{ \overset{\PP}{\underset{K \to \infty}{\to}} }

\newcommand{\toK}{ \overset{}{\underset{K \to \infty}{\to}} }	
\newcommand{\simK}{ \overset{}{\underset{K \to \infty}{\sim}} }	
\newacro{TDMA}{Time Division Multiple Access}

\newacro{ProSe}{Proximity Service}

\newacro{eMBMS}{enhanced Multimedia Broadcast Multicast Service}

\newacro{H-ARQ}{Hybrid Automatic Repeat reQuest}

\newacro{i.i.d.}{independent and identically distributed}

\newacro{AWGN}{Additive White Gaussian Noise}

\newacro{BC}{Broadcast Channel}

\newacro{D2D}{Device-to-Device}

\newacro{VSC}{Virtual Small Cell}

\newacro{LSAS}{Large Scale Antenna System}

\newacro{SE}{Spectral Efficiency}

\newacro{SISO}{Single-Input Single-Output}

\newacro{SIMO}{Single-Input Multiple-Output}

\newacro{MISO}{Multiple-Input Single-Output}

\newacro{MIMO}{Multiple-Input Multiple-Output}

\newacro{SNR}{Signal to Noise Ratio}

\newacro{SINR}{Signal to Interference Noise Ratio}

\newacro{EE}{Energy Efficiency}

\newacro{LOS}{Line of Sight}

\newacro{NLOS}{Non-Line of Sight}

\newacro{MCS}{Modulation and Coding Scheme}

\newacro{PA}{Power Amplifier}

\newacro{ICT}{Information and Communication Technology}

\newacro{TDD}{Time Division Duplixing}

\newacro{(f)eICIC}{(further) enhanced Inter-Cell Interference}

\newacro{ABS}{Almost Blank Subframes}

\newacro{CRE}{Cell Range Expansion}

\newacro{CoMP}{Coordinated Multi-Point}

\newacro{ePDCCH}{enhanced Physical Downlink Control Channel}

\newacro{HetNet}{Heterogeneous Network}

\newacro{JSDM}{Joint Spatial Division Multiplexing}

\newacro{CCM}{channel correlation matrice}

\newacro{NTCQ}{Noncoherent Trellis-Coded Quantization}

\newacro{FDD}{Frequency Division Duplexing}

\newacro{CSI}{Channel State Information}

\newacro{CSIT}{Channel State Information at the Transmitter}

\newacro{CSIR}{Channel State Information at the Receiver}

\newacro{IMEC}{Interuniversity Microelectronics Centre}

\newacro{PCI}{Physical Cell ID}

\newacro{CRS}{Common Reference Signal}

\newacro{TP}{Transmission Point}

\newacro{RRH}{Remote Radio Head}

\newacro{LTE-A}{Long Term Evolution-Advanced}

\newacro{RF}{Radio Frequency}

\newacro{MU}{Multi User}

\newacro{MUBF}{Multi User Beamforming}

\newacro{4G}{Fourth-Generation}

\newacro{5G}{Fifth-Generation}

\newacro{TUDR}{Typical User Data Rate}

\newacro{KPI}{Key Performance Indicator}

\newacro{RS}{Rate Splitting}

\newacro{HRS}{Hierarchical Rate Splitting}

\newacro{3GPP}{3rd Generation Partnership Project}

\newacro{AI}{Artificial Intelligence}

\newacro{ARM}{Available Resource Map}

\newacro{BS}{Base Station}

\newacro{CAPEX}{Capital Expenditure}

\newacro{CDF}{Cumulative Distribution Function}

\newacro{CR}{Cognitive Radio}

\newacro{DL}{Down Link}

\newacro{GPS}{Global Positioning System}

\newacro{IC}{Interference Cartography}

\newacro{IEEE}{Institute of Electrical and Electronics Engineers}

\newacro{LTE}{Long Term Evolution}

\newacro{MAE}{Mean Absolute Error}

\newacro{MCD}{Measurement Capable Device}

\newacro{MDT}{Minimization of Drive Tests}

\newacro{MSE}{Mean Squared Error}

\newacro{MMSE}{Minimum Mean Squared Error}

\newacro{PL}{Path Loss}

\newacro{PU}{Primary User}

\newacro{REM}{Radio Environment Map}

\newacro{RMSE}{Root Mean Square Error}

\newacro{RRM}{Radio Resource Management}

\newacro{RSCP}{Received Signal Code Power}

\newacro{RSRP}{Reference Signal Received Power}

\newacro{SON}{Self-Organized Network}

\newacro{OeM}[O\&M]{Operation and Maintenance}

\newacro{OFDM}{Orthogonal Frequency Division Multiplexing}

\newacro{OMC}{Operations and Maintenance Centre}

\newacro{OPEX}{Operational Expenditure}

\newacro{QoS}{Quality of Service}

\newacro{TCE}{Trace Collection Entity}

\newacro{UE}{User Equipment}

\newacro{UTRAN}{Universal Terrestrial Radio Access Network}

\newacro{UWAN}{Unlicensed Wide Area Network}

\newacro{WRAN}{Wireless Regional Area Network}

\newacro{MBF}{Matched Beamforming}

\newacro{BF}{Beamforming}

\newacro{DoF}{Degree of Freedom}

\newacro{GKA}{Genetic K-means Algorithm}

\newacro{ISD}{Inter Site Distance}

\newacro{RAN}{Radio Access Network}

\newacro{ZF}{Zero Forcing}

\newacro{RZF}{Regularized Zero Forcing}

\newacro{TTP}{Two-Tier Precoder}

\newacro{2D}{two-dimensional}

\newacro{3D}{three-dimensional}

\newacro{eNB}{Evolved NodeB}

\newacro{PPP}{Poisson Point Process}

\newacro{KL}{Karhunen-Loeve}

\newacro{SQ}{Scalar Quantization}

\newacro{VQ}{Vector Quantization}

\newacro{CMI}{Conditional Mutual Information}

\newtheorem{proposition}{\textbf{Proposition}}
\newtheorem{theorem}{\textbf{Theorem}} 

\newtheorem{corollary}[theorem]{\textbf{Corollary}}

\newtheorem{assumption}{\textbf{Assumption}}

\DeclareGraphicsExtensions{.eps}

\graphicspath{{.//}}	

\renewcommand{\contentsname}{}
\renewcommand{\listfigurename}{}

\begin{document}
\title{Device-to-Device Aided Multicasting}



 \author{%
   \IEEEauthorblockN{Thomas Varela Santana\IEEEauthorrefmark{1}\IEEEauthorrefmark{2},
                     Richard Combes\IEEEauthorrefmark{2}
                     and Mari Kobayashi\IEEEauthorrefmark{2}
                     }
   \IEEEauthorblockA{\IEEEauthorrefmark{1}%
                     Orange Labs Networks, Ch\^{a}tillon, France,
                     Email: thomas.varelasantana@orange.com}
   \IEEEauthorblockA{\IEEEauthorrefmark{2}%
                     Laboratoire des Signaux et Systèmes (L2S), CentraleSup\'{e}lec-CNRS-Universit\'{e} Paris-Sud, Gif-sur-Yvette, France,\\
                     Email: \{richard.combes, mari.kobayashi\}@centralesupelec.fr}
 }

\maketitle

\begin{abstract}
	We consider a device-to-device (D2D) aided multicast channel, where a transmitter wishes to convey a common message to many receivers and these receivers cooperate with each other. We propose a simple computationally efficient scheme requiring only statistical channel knowledge at transmitter. Our analysis in general topologies reveals that, when the number of receivers $K$ grows to infinity, the proposed scheme guarantees a multicast rate of ${1 \over 2} \log_2(1 + \beta \ln K )$ with high probability for any $\beta < \beta^\star$ where $\beta^\star$ depends on the network topology. This scheme undergoes a phase transition at threshold $\beta^\star \ln K$ where transmissions are successful/unsuccessful with high probability when the SNR is above/below this threshold. We also analyze the outage rate of the proposed scheme in the same setting.  
\end{abstract}


\section{Introduction}\label{sec:introduction}


We consider the multicast channel, where a single transmitter sends a common message to many receivers in the presence of fading~\cite{lopez2004multiplexing,gopala2005throughput,sidiropoulos2006transmit}. Since the common message must be decoded by all receivers, 
the multicast capacity of the fading broadcast channel is limited by the worst user. 
For the case of the i.i.d. Rayleigh fading channel, it is known that the multicast capacity vanishes inversely to the number $K$ of users as $K$ grows ~\cite{jindal2006capacity}. 
Despite its vanishing rate, the multicast channel is relevant to two scenarios in wireless crowded networks.
The first scenario is the wireless edge caching. It has been shown that the traffic during the peak hours can be significantly reduced by caching popular contents during off-peak hours, and delivering these popular contents using multicasting~(see e.g. \cite{maddah2014fundamental,paschos2016wireless} and references therein). A novel user selection scheme requiring only statistical channel knowledge has been proposed recently in~\cite{ngo2018scalable} for the case of i.i.d. Rayleigh fading channel.  The second scenario is so-called enhanced Multimedia Broadcast Multicast Service (eMBMS) aided by Proximity Service (ProSe), standardized by the \ac{3GPP}. These services reduce the network load by multicasting common data to public safety devices~\cite{nokiaglobal} which can use \acf{D2D} communication~\cite{3GPP_D2D}. 
In this work, motivated by these two scenarios, we study D2D-aided multicasting without \ac{CSIT} to overcome the vanishing effect of the multicast rate. We consider a general network topology where the transmitter has only statistical channel knowledge. In this setup, we address the fundamental question: \textit{can D2D without CSIT increase the achievable multicasting rate ?} 

We propose a two-stage transmission scheme assuming no \ac{CSIT}: in the first stage the transmitter sends a common message, while in the second stage, the subset of users whom have successfully decoded retransmit this information simultaneously by using the same codeword. We study this two-stage scheme for two metrics of interest:
\begin{itemize}
	\item The average multicast rate, which is the expected number of successfully decoded bits per channel use achieved by a user chosen uniformly at random;
	\item The outage rate, which is the maximum rate at which all the users can decode the message with an error probability of $\epsilon$.
\end{itemize}
We show that by carefully choosing the transmission rate, we can answer positively to our main question. More specifically, our contributions is two-fold and summarized below:
\begin{itemize}
	\item[1)] We identify conditions on the network topology for which the average multicast rate grows with the number of users $K$ and the outage rate does not vanish with $K$.
	\item[2)] We provide tractable, asymptotic expressions for both the multicast rate and the outage rate in the regime of a large number of users.
\end{itemize}
A similar two-phase scheme has been studied in~\cite{khisti2006fundamental}. However this work is different from ours in its assumption and concept. First, the 
channel statistic of all users is assumed to be symmetric in \cite{khisti2006fundamental}. In fact, this is a special case of our model corresponding to a single class $C=1$ (see subsection \ref{subsec:matrices}). Second, the metric of the work \cite{khisti2006fundamental} is on the error probability decay by exploiting channel hardening via space-time code, while we aim at achieving a scalable multicast rate by exploiting multiuser diversity.

The rest of the paper is organized as follows. Section~\ref{sec:model} presents the system model. Sections~\ref{sec:multicast_rate} and~\ref{sec:outage_rate} studies the multicast and outage rate metrics respectively. Numerical experiments are provided in Section~\ref{sec:numerical}. Section~\ref{sec:conclusion} concludes the paper. We denote by $\ln$ the natural logarithm and $\log_2$ the binary logarithm.	
\section{Model}\label{sec:model}
\subsection{Channel Model}\label{subsec:channel_model}

We consider a multicast channel, where a station (indexed by $0$) wants to convey the same message to $K$ users indexed by $i=1,\dots,K$. Time is slotted, and in each time slot, $y_i$ the received signal by $i=1,\dots,K$ is given by:
$$
	y_i =  \sum_{j=0}^K x_j h_{j,i} + n_i,
$$
with $(x_i)_{i=0,\dots,K}$ the signal transmitted by $i=0,\dots,K$, $\mathbf{h} = (h_{i,j})_{i,j=0,\dots,K}$ are the channel coefficients and $n_i \sim {\cal N}(0,1)$ is Additive White Gaussian Noise (AWGN).  The channel coefficients $\mathbf{h}$ are assumed to be independent, Gaussian complex random variables with mean $0$ and variance $\gamma_{i,j} = \EE(|h_{i,j}|^2)$. The transmitted signals have unit power $\EE(|x_i|^2) \le 1$. This is without loss of generality, otherwise simply replace $x_i$ by ${x_i \over \sqrt{\EE(|x_i|^2)}}$ and  $h_{i,j}$ by $\sqrt{\EE(|x_i|^2)} h_{i,j}$. 

Matrix $\Gamma = (\gamma_{i,j})_{i,j=0,\dots,K}$ represents the channel statistics, and captures the topology of the network. Channel coefficients during successive time slots are assumed to be independent. No channel state information is available at the transmitter: the transmission strategy should only depend on the \emph{statistics} of the channel (i.e. $\Gamma$), but not on the channel coefficients $\mathbf{h}$.
\subsection{Proposed scheme}\label{subsec:objectives}
Throughout the article we will study the following transmission scheme, which utilizes two time slots. Let $\mathbf{h}$, $\mathbf{h}'$, the channel coefficients during the two successive time slots.
\begin{itemize}
\item During the first time slot, the station broadcasts a message at rate $\log_2(1 + s)$, where $s$ is a parameter of the scheme and can be chosen depending on the channel statistics $\Gamma$.
\item At the end of the first time slot, users attempt to decode, and user $i$ decodes successfully if and only if the received Signal to Noise Ratio (SNR) is greater than $s$, i.e. $|h_{0,i}|^2 \ge s$ (\footnote{This is valid if we assume that time slots are long enough so that one may use an efficient code for the AWGN channel.})
\item All users that have successfully decoded the message in the first slot retransmit this message in the second slot. As before, user $j$ decodes successfully at the end of the second time slot if the SNR is above $s$, i.e. $|\sum_{j=1}^{K} Z_j(s) h_{j,i}'|^2 \ge s$,
where $Z_1(s),\dots,Z_K(s)$ are binary variables with $Z_j(s)=1$ if $j$ has decoded successfully during the first time slot and $0$ otherwise.
\end{itemize}
\subsection{Performance Measures}
	We say that user $i$ decodes if and only if he successfully decodes either in the first or second slot. Denote by $P_i(s)$ the probability that $i=1,\dots,K$ decodes, $\bar{P}(s) ={1 \over K} \sum_{i=1}^K P_i(s)$ the probability that a user chosen uniformly at random amongst $i=1,\dots,K$ decodes, and $P_{+}(s)$ probability that all users decode. We study two performance measures for our problems: the multicast rate:
$$
	R^{m} = {1 \over 2} \max_{s \ge 0} \{ \log_2(1 + s) \bar{P}(s) \},
$$
which is the expected number of bits received by a user chosen uniformly at random per time slot, and the outage rate:
$$
	R^{o} = {1 \over 2} \log_2(1 + s) \text{ with } s \text{ solution to } P_{+}(s) = 1-\epsilon.
$$
which is the largest rate such that \emph{all} users decode with probability at least $1-\epsilon$, with $\epsilon$ some fixed reliability level. Both performance measures are interesting for different scenarios: $R^m$ seems more appropriate when considering say video streaming where the goal is that on average most users receive enough information, whereas $R^o$ seems more suited to applications such as broadcasting of safety information, where it is important that \emph{all} users obtain the message.
\subsection{The Block Model}\label{subsec:matrices}
To study large systems, we introduce a block model for $\Gamma$. Users $1,\dots,K$ are partitioned in classes $1,\dots,C$, where $c_i \in \{1,\dots,C\}$ indicates the class of user $i$, and $c_0 = 0$ by convention. There are $K \alpha_c$ users of class $c$, with $\alpha_c > 0$ the proportion of users of class $c$ and $\sum_{c=1}^C \alpha_c = 1$. Matrix $\Gamma$ is a block matrix, so that the mean channel gains between two users depends solely on their class. Namely, $\gamma_{i,j} = g_{c_i,c_j}$ for all $i,j=0,\dots,K$. Define $G = \max_{c,c'=0,\dots,C} g_{c,c'}$ the largest entry of $\Gamma$, so that $\max_{i,j=0,\dots,K} \gamma_{i,j} = G$. This model can represent any network topology, as the number of classes may be arbitrary. We introduce the following natural assumption.
\medskip
\begin{assumption}\label{ass:connect}
	Any class of users can be reached in two transmissions, so that for all $c=1,\dots,C$ there exists $c'$ such that $g_{0,c'} g_{c',c} > 0$.
\end{assumption}
\medskip
\subsection{Baseline}
We will compare the performance of this scheme to the most natural baseline which is the same scheme where only the first time slot is used. The performance of the baseline is recalled below. 
\medskip
\begin{proposition}
	The performance of the baseline scheme is:
	\begin{align*}
		R^m &= \max_{s \ge 0} \{ \log_2(1+s) \sum_{c=1}^C \alpha_c e^{-{s \over g_{0,c}}} \}  \underset{K \to \infty}{=} {\cal O}(1), \\
		R^o &= \log_2\Big( 1 + {1 \over K} \ln\Big({1 \over 1 - \epsilon}\Big)  \Big(\sum_{c=1}^C {\alpha_c \over g_{0,c}}\Big)^{-1} \Big) \underset{K \to \infty}{=} {\cal O}\Big({1 \over K}\Big).
	\end{align*}
\end{proposition}
\bp
	In the baseline scheme, user $i$ successfully decodes if and only if $|h_{0,i}|^2 \ge s$. Since $|h_{0,i}|^2$ follows an exponential distribution with mean $\gamma_{0,i}$, we have:
	\begin{align*}
		P_i(s) = \PP(|h_{0,i}|^2 \ge s) = e^{- {s \over \gamma_{0,i}}},
	\end{align*}
	and averaging:
	\begin{align*}
		\bar{P}(s) = {1 \over K} \sum_{i=1}^K  e^{- {s \over \gamma_{0,i}}} = \sum_{c=1}^C \alpha_c e^{-{s \over g_{0,c}}}.
	\end{align*}
	which is the announced result. 
	
	Once again $i$ decodes successfully if and only if $|h_{0,i}|^2 \ge s$,  $|h_{0,i}|^2$ follows an exponential distribution with mean $\gamma_{0,i}$ and $h_{0,1},\dots,h_{0,K}$ are independent so that:
	\begin{align*}
		P_+(s) &= \PP( |h_{0,i}|^2 \ge s \,,\, i=1,\dots,K)  
		= \prod_{i=1}^K \PP( |h_{0,i}|^2 \ge s) \\
		&= \exp\left\{ -s \sum_{i=1}^K {1 \over \gamma_{0,i}} \right\} 
		= \exp\left\{ -s K \sum_{c=1}^C   {\alpha_c \over g_{0,c}} \right\}.
	\end{align*}
	The outage rate is $\log_{2}(1+s)$ with $s$ solution to the equation $P_{+}(s) = 1-\epsilon$, so replacing yields the announced result.
\ep
		   	
\section{Multicast Rate}\label{sec:multicast_rate}
In this section we study the multicast rate of the proposed scheme, and show that it undergoes a phase transition in the regime of a large number of users $K \to \infty$, so that the probability of success $\bar{P}(s)$ becomes constant-by-parts and may be computed explicitly as a function of $g$ and $\alpha$.
\subsection{Success probability}
We first prove Proposition~\ref{prop:multicast_delivery_proba}, a formula for the probability of success $P_i(s)$, which will serve as the backbone of our analysis. This result shows that the success probability $P_i(s)$ can be controlled by examining the fluctuations of $X_i(s)$, a sum of independent Bernoulli random variables.
\medskip
\begin{proposition}\label{prop:multicast_delivery_proba}
	For any $i=1,\dots,K$ and $s \ge 0$ we have:
	$$
		P_i(s) = 1 - (1-e^{- {s \over \gamma_{0,i}}})\EE\left(1 - \exp\left\{- {s \over X_i(s)}\right\}\right),
	$$
	where $Z_1(s),\dots,Z_K(s)$ are independent random variables in $\{0,1\}$ with $\EE(Z_i(s)) = e^{- {s\over \gamma_{0,i}}}$ and $X_i(s) = \sum_{j=1}^K Z_j(s) \gamma_{j,i}$.
\end{proposition}
\medskip
\bp
	Consider $\mathbf{h} = (h_{i,j})_{i,j=0,\dots,K}$ and $\mathbf{h}' =(h_{i,j}')_{i,j=0,\dots,K}$ the channel coefficients during the first and second time slot respectively. By assumption $\mathbf{h}$ is independent from $\mathbf{h}'$. Denote by $\mathbf{Z}(s) = (Z_1(s),\dots,Z_K(s))$ the outcome of the first time slot, where $Z_i(s) = 1$ if user $i$ decodes correctly at the first time slot. We may write $Z_i(s) = \indic\{ |h_{0,i}|^2 \ge s \}$. Since $|h_{0,i}|^2$ has exponential distribution with mean $\gamma_{0,i}$, it follows that $\EE(Z_i(s)) = e^{-{s \over \gamma_{0,i}}}$. Furthermore, since $h_{0,1},\dots,h_{0,K}$ are independent, so are  $Z_1(s),\dots,Z_K(s)$. 
	
	Consider user $i$. Conditionally to the value of $\mathbf{Z}(s)$, user $i$ does not decode successfully in the first phase if and only if $Z_i(s) = 0$. If $Z_i(s) = 0$, he does not decode successfully in the second phase if and only if $|\sum_{j \ne i} h_{j,i}' Z_j(s)|^2 = |\sum_{j=1}^K h_{j,i}' Z_j(s)|^2 \le s$, where $|\sum_{j=1}^K h_{j,i} Z_j(s) |^2$ has exponential distribution with mean $\sum_{j=1}^K \gamma_{j,i} Z_j(s) = X_i(s)$, since $\mathbf{Z}(s)$ is independent of $\mathbf{h}'$. Hence 
	$$
	\PP( i \text{ does not decode}  | \mathbf{Z}(s) )= (1-Z_i(s))(1 - e^{-{s \over X_i(s)}}).
	$$ 
	Taking expectations over $\mathbf{Z}(s)$ yields the result \ep.

\subsection{Asymptotic behavior}
We now analyze how the multicast rate scales in the regime of a large number of users. Define:
	$$
	\beta_c = \max_{c'=1,\dots,C}\{  g_{0,c'} \indic\{ g_{c',c} > 0 \} \},
	$$ 
the largest value of $g_{0,c'}$, the mean channel gain from the station to class $c'$, where class $c'$ can communicate with class $c$ in the second phase i.e. $g_{c',c} > 0$. Further define the minimum value $\beta^\star = \min_{c=1,\dots,C} \beta_c$.  Theorem~\ref{th:main_result_multicast} shows that in the limit of a large number of users, for any user $i$, the success probability $P_i$ undergoes a phase transition at the value $\beta_{c_i} \ln K$. Namely transmissions are always successful above this threshold, and always unsuccessful below. This has 3 consequences:
\begin{itemize}
\item Our scheme transmits at rate ${1 \over 2} \log_2(1 + \beta \ln K)$ with an arbitrarily high probability of success for any $\beta < \beta^\star$. 
\item As $K \to \infty$, the multicast rate of our scheme scales as ${\cal O}( \ln \ln K)$ while the baseline yields a multicast rate of ${\cal O}(1)$. So considering two slots instead of one has a dramatic impact on performance. Considering more than two time slots does not seem to improve this scaling.
\item To obtain an order-optimal rate, one can set $s = \beta \ln K$ for any $\beta < \beta^\star$, so that optimizing over $s$ is not needed.
\end{itemize}
\medskip
\begin{theorem}\label{th:main_result_multicast}
	(i) For any $\beta > 0$ and $i=1,\dots,K$ we have:
	$$
		P_i\left( \beta \ln K \right) \toK \begin{cases}  1 & \text{ if } \beta < \beta_{c_i} \\  0 & \text{otherwise.} \end{cases}
	$$
	(ii) We have:
	$$
		\bar{P} \left( \beta \ln K \right) \toK  \sum_{c=1}^C \alpha_c \indic\{\beta < \beta_c\},  
	$$
\end{theorem}
\medskip
\bp
	From Proposition~\ref{prop:multicast_delivery_proba}, we have:
	$$
		P_i(\beta \ln K) = 1 - (1 - K^{- {\beta \over g_{0,c_i}}})\EE(1 - e^{- {\beta \ln K \over X_i(\beta \ln K)}}).
	$$
	We have $K^{- {\beta \over g_{0,c_i}}} \toK 0$ so that:
	$$
		\limK P_i(\beta \ln K) = \limK \EE(e^{- {\beta \ln K \over X_i(\beta \ln K)}}).
	$$
	\underline{The $\beta > \beta_{c_i}$ case.} Assume that $\beta > \beta_{c_i}$, and we control the expectation of $X_i(\beta \ln K)$. Since $\EE(Z_j(s)) = e^{-{s \over g_{0,c_j}}}$:
	\begin{align*}
		\EE\Big({X_i(\beta \ln K) \over \beta \ln K}\Big) &= \sum_{j=1}^K { \gamma_{j,i}  K^{-{\beta \over \gamma_{0,j}}} \over \beta \ln K} 
		 																					= \sum_{c=1}^C {\alpha_c g_{c,c_i} K^{1 - {\beta \over g_{0,c}} } \over \beta \ln K} \\
		 																					&\le  \sum_{c=1}^C {\alpha_c g_{c,c_i} K^{1 - {\beta \over \beta_{c_i}}} \over \beta \ln K} \toK 0
	\end{align*}
	since for all $c$ either $g_{c,c_i}= 0$ or $g_{0,c} \le \beta_{c_i}$, and $\beta > \beta_{c_i}$. Therefore, ${X_i(\beta \ln K) \over \beta \ln K}$ converges to $0$ in $L^{1}$ so that it converges to $0$ in distribution as well. Since $x \mapsto e^{-{1 \over x}}$ is both continuous and bounded we get:
	$$
		\limK P_i(\beta \ln K) = \limK \EE(e^{- {\beta \ln K \over X_i(\beta \ln K)}}) = 0.
	$$	
	
	\underline{The $\beta < \beta_{c_i}$ case.} Now consider $\beta < \beta_{c_i}$. We control the moments of $X_j(s)$:
	$$
	 \EE(X_i(s)) = \sum_{j=1}^K \gamma_{j,i} e^{-{s \over \gamma_{0,j}}},
	$$
and since $Z_1(s),\dots,Z_K(s)$ are independent:
	\begin{align*}
		\var(X_i(s)) &= \sum_{j=1}^K \gamma_{j,i}^2 e^{-{s \over \gamma_{0,j}}}(1- e^{-{s \over \gamma_{0,j}}} ) \\
									&\le G \sum_{j=1}^K \gamma_{j,i} e^{-{s \over \gamma_{0,j}}}  = G  \EE(X_i(s)).
	\end{align*}
	Apply Chebychev's inequality:
	\begin{align*}
		\PP\left( X_i(s) \le  {\EE(X_i(s)) \over 2}\right) 
		&\le \PP\left( |X_i(s) - \EE(X_i(s)) | \ge {\EE(X_i(s)) \over 2}\right) \\
																	&\le {4 \var(X_i(s)) \over \EE(X_i(s))^2}
																	\le {4 G \over \EE(X_i(s))}  
	\end{align*}
	using the previous bound. Hence
	\begin{equation}\label{eq:chebychev}
		\PP\left( X_i(s) \ge  {\EE(X_i(s)) \over 2}\right) \ge 1 - {4 G \over \EE(X_i(s))}.
	\end{equation}
	Since $x \mapsto e^{-{1 \over x}}$ is increasing:
	\begin{align*}
		\EE( & e^{- {\beta \ln K \over X_i(\beta \ln K)}}) \\
		&\ge  \PP\left(X_i(\beta \ln K) \ge {\EE(X_i(\beta \ln K)) \over 2} \right) e^{- {2 \beta \ln K \over \EE(X_i(\beta \ln K))}}.
	\end{align*}
	Consider ${\hat c}$ such that $g_{{\hat c},c_i} > 0$ and $g_{0,{\hat c}} = \beta_{c_i}$. Then:
	\begin{align*}
		\EE\left({X_i(\beta \ln K) \over \beta \ln K}\right) &=  \sum_{c=1}^C {\alpha_c g_{c,c_i} K^{1 - {\beta \over g_{0,c}} } \over \beta \ln K} \\
		&\ge {\alpha_{{\hat c}} g_{{\hat c},c_i} K^{1 - {\beta \over \beta_{c_i}}} \over \beta \ln K} \toK \infty. 
	\end{align*}
	Replacing in~\eqref{eq:chebychev} we deduce 
	\begin{align*}
	\PP\left(X_i(\beta \ln K) \ge {\EE(X_i(\beta \ln K)) \over 2} \right) \toK 1,
	\end{align*}
	and $e^{- {2 \beta \ln K \over \EE(X_i(\beta \ln K))}} \toK 1$ so that:
	\begin{align*}
			\limK P_i(\beta \ln K) = \limK \EE(e^{- {\beta \ln K \over X_i(\beta \ln K)}}) = 1.
	\end{align*}
	which completes the proof of statement (i). Statement (ii) follows from the fact that $\bar{P}(s) = {1 \over K} \sum_{i=1}^K P_i(s)$.
\ep
\medskip
\begin{corollary}
		For any $\beta < \beta^\star$ we have:
		$$
			R^m \ge {1 - o(1) \over 2} \log_2(1 + \beta \ln K) \,,\, K \to \infty.
		$$
\end{corollary}
\medskip
\subsection{Non asymptotic behavior}
	We may state Theorem~\ref{th:main_res_multicast_approx}, a further result which gives a tractable, accurate approximation for $1 - P_j(\beta \ln K)$ in the regime where $P_j(\beta \ln K) \toK 1$ i.e. whenever $\beta < \beta_{c_i}$. Two main facts are worth mentioning:
\begin{itemize}
\item This approximation is very accurate even for modest size systems (say $K \ge 50$) as shown by our numerical experiments (see section~\ref{sec:numerical}).
\item Due to it's accuracy, it allows to find the optimal value of $s$ given $g$ and $\alpha$ in a tractable manner, so that finite size systems can be dealt with efficiently.
\end{itemize}
\medskip
\begin{theorem}\label{th:main_res_multicast_approx}
	Consider $\beta < \beta_c$, then:
	$$
			1 - P_i\left( \beta \ln K \right) \simK 1 - \exp\left\{ { \beta \ln K \over \sum_{c=1}^C \alpha_c g_{c,c_i} K^{1 - {\beta \over g_{0,c}}}} \right\}.
	$$
\end{theorem}
\bp
Define $f(x) = 1 - e^{-{1 \over x}}$. Throughout the proof consider $i$ fixed, and define $V =  { X_i(\beta \ln K) \over \beta \ln K}$.

From Proposition~\ref{prop:multicast_delivery_proba}:
\begin{align*}
	1 - P_i\left( \beta \ln K \right) = (1 - K^{-{\beta \over \gamma_{0,i}}})\EE(f(V)) \simK \EE(f(V)).
\end{align*}
since $K^{-{\beta \over \gamma_{0,i}}} \toK 0$. Define:
$$
	v_K = \EE(V) =  {\sum_{c=1}^C \alpha_c g_{c,c_i} K^{1 - { \beta \over g_{0,c}}} \over  \beta \ln K} \toK \infty.
$$
since $\beta < \beta^\star$. To complete the proof, it suffices to show that 
$$
	\EE(f(V)) \simK f(v_K).
$$
Now consider $\delta \in (0,1)$ fixed and let us bound $\EE(f(V))$.

\underline{Upper bound:} Since $x \mapsto f(x)$ is decreasing:
\begin{align*}
	f(V) &=  f(V) \indic\{V \le (1-\delta) v_K \}  + f(V)\indic\{V \ge (1-\delta) v_K \} \\
		 &\le \indic\{V \le (1-\delta) v_K \} + f((1-\delta)v_K).
\end{align*}
So taking expectations and using Chernoff's inequality, which is recalled as proposition~\ref{prop:chernoff} in appendix:
\begin{align*}
	\EE(f(V)) &\le \PP(V \le (1-\delta) v_K) + f((1-\delta)v_K) \\
				&\le e^{-{\delta^2 v_K \over 3} } + f((1-\delta)v_K).
\end{align*}
Now using the facts $v_K \toK \infty$  so that $f(v_K) \simK {1 \over v_K}$ and $v_K e^{-{\delta^2 v_K \over 3} } \toK 0$ we get:
\begin{align*}
	\lim\sup_{K \to \infty} {\EE(f(V)) \over f(v_K)} \le {1 \over 1-\delta}.
\end{align*}

\underline{Lower bound:} Similarly:
\begin{align*}
	f(V) &=  f(V) \indic\{V \le (1+\delta) v_K \}  + f(V)\indic\{V \ge (1+\delta) v_K \} \\
	&\ge   \indic\{V \le (1+\delta) v_K \} f((1+\delta)v_K)
\end{align*}
So taking expectations and using Chernoff's inequality, which is recalled as proposition~\ref{prop:chernoff} in appendix:
\begin{align*}
	f(V) &\ge \PP(V \le (1+\delta) v_K) f((1+\delta)v_K)\\ 
	&\ge  (1 - e^{-{\delta^2 v_K \over 3}}) f((1+\delta)v_K).
\end{align*}
Using the facts $v_K \toK \infty$ so that $f(v_K) \simK {1 \over v_K}$ and $e^{-{\delta^2 v_K \over 3} } \toK 0$ we get:
\begin{align*}
	\lim\inf_{K \to \infty} {\EE(f(V)) \over f(v_K)} \ge {1 \over 1+\delta}.
\end{align*}

\underline{Putting it together:} The above holds for any $\delta \in (0,1)$. So we have proven that $\delta$ arbitrairly small:
\begin{align*}
	{1 \over 1+\delta} \le \lim\inf_{K \to \infty} {\EE(f(V)) \over f(v_K)} \le \lim\sup_{K \to \infty} {\EE(f(V)) \over f(v_K)} \le {1 \over 1-\delta}.
\end{align*}
Hence we have proven
\begin{align*}
	\EE(f(V)) \simK f(v_K).
\end{align*}
which concludes the proof.
\ep

\section{Outage Rate}\label{sec:outage_rate}
We now turn to the outage rate, and we show that in the regime of a large number of users, the outage rate may be computed explicitly as a function of $g$ and $\alpha$. We further show that, while the outage rate of the baseline scheme vanishes when $K \to \infty$ our scheme guarantees a constant outage rate. 
\subsection{Success probability}
As in the multicast case, we express the outage rate as a function of $X_1(s),\dots,X_K(s)$ in Proposition~\ref{prop:outage_delivery_proba}.
\medskip
\begin{proposition}\label{prop:outage_delivery_proba}
	For any $s \ge 0$ we have:
	\begin{align*}
	P_{+}(s) = \EE \left[\exp\left\{ - s \sum_{i=1}^K {1-Z_i(s) \over X_i(s)}\right\}\right],
	\end{align*}
	where $Z_1(s),\dots,Z_K(s)$ are independent random variables in $\{0,1\}$ and $\EE(Z_i(s)) = e^{- {s\over \gamma_{0,i}}}$ for $i=1,\dots,K$ and $X_i(s) = \sum_{j=1}^K Z_j(s) \gamma_{j,i}$.
\end{proposition}
\medskip
\bp We use the same notation as in the proof of Proposition~\ref{prop:multicast_delivery_proba}. User $i$ successfully decodes if and only if either $Z_i(s) = 1$ or $|\sum_{j=1}^K Z_j(s) h'_{j,i} |^2 \ge s$. So $i$ decodes if and only if:
$$\left|\sum_{j=1}^K Z_j(s) h'_{j,i} \right|^2 \ge s(1-Z_i(s)).
$$
Now:
\begin{align*}
	\PP( & \text{all decode} | \mathbf{Z}(s)) \\ 
	&= \PP\Big( \Big| \sum_{j=1}^K Z_j(s) h'_{j,i} \Big|^2 \ge s(1-Z_i(s)) \, , \,\forall i | \mathbf{Z}(s)\Big) \\
	&= \prod_{i=1}^K \PP\Big( \Big| \sum_{j=1}^K Z_j(s) h'_{j,i} \Big|^2 \ge s(1-Z_i(s)) | \mathbf{Z}(s) \Big) \\
 	&= \exp \left\{-s \sum_{i=1}^K {1-Z_i(s) \over X_i(s)}\right\}.
\end{align*}
since conditional to $\mathbf{Z}(s)$, $|\sum_{j=1}^K Z_j(s) h'_{j,i}|^2$ has exponential distribution with mean $X_i(s)$, and $\mathbf{h}'$ has independent entries. Averaging over $\mathbf{Z}(s)$ yields the result.
\ep

\subsection{Asymptotic behavior}
We now prove our main result concerning outage rate stated in Theorem~\ref{th:main_result_outage}. The main proof element is to show that, for any $s$ random variable $\sum_{i=1}^K {1-Z_i(s) \over X_i(s)}$ concentrates around its expectation when the number of users $K$ grows large. The following facts should be noted:
\begin{itemize}
\item In the regime of a large number of users, the outage rate converges to a non-zero finite value. Recall that the baseline scheme only guarantees a vanishing outage rate. Hence our scheme is very efficient in countering the variability of the channel coefficients and induces a perfect channel hardening.
\item The outage rate can be computed explicitly as a function of $g$ and $\alpha$ (see corollary~\ref{cor:main_result_outage}) by a zero-finding method such as  bisection or Newton-Raphson.
\item In fact, this asymptotic result provides a very good approximation even for systems of modest size (say $K \ge 10$), as shown by numerical experiments (see  section~\ref{sec:numerical}).
\end{itemize}
\medskip
\begin{theorem}\label{th:main_result_outage}
	For any $s \ge 0$ we have:
	\begin{align*}
		P_{+}(s) \toK \exp\left\{ - s \sum_{c=1}^C {\alpha_c(1 - e^{-{s \over g_{0,c}}}) \over \sum_{c'=1}^C \alpha_{c'} g_{c',c} e^{- {s \over g_{0,c'}}} }\right\}.
	\end{align*}
\end{theorem}
\medskip
\bp
	We first prove the following fact:
	\begin{align*}
		\sum_{i=1}^K { 1 - Z_i(s) \over X_i(s)}  \PtoK \sum_{i=1}^K {1 - e^{- {s \over \gamma_{0,i}}} \over \EE(X_i(s))}.
	\end{align*}
	We bound the error as follows:
\begin{align*}
	& \left| \sum_{i=1}^K { 1 - Z_i(s) \over X_i(s)}   - \sum_{i=1}^K {1 - e^{- {s \over \gamma_{0,i}}} \over \EE(X_i(s))} \right| \\
	& \le \left| \sum_{i=1}^K {1-Z_i(s) \over X_i(s)} - {1-Z_i(s) \over \EE(X_i(s))} \right| 
	+ \left| \sum_{i=1}^K {Z_i(s)- e^{- {s \over \gamma_{0,i} }} \over \EE(X_i(s))} \right|.
\end{align*}
We now prove that both terms go to $0$ in probability.

\underline{First term} We have: 
\begin{align*}
	 \left| \sum_{i=1}^K {1-Z_i(s) \over X_i(s)} - {1-Z_i(s) \over \EE(X_i(s))} \right| 
	 &\le  \sum_{i=1}^K \left| {1 \over X_i(s)} - {1 \over \EE(X_i(s))} \right| \\
	 &=  \sum_{i=1}^K {|X_i(s) - \EE(X_i(s))| \over X_i(s) \EE(X_i(s))}.
\end{align*}
Furthermore
$$
\EE(X_i(s)) = K \sum_{c=1}^C \alpha_c g_{c,c_i} e^{- {s \over g_{0,c}}}  \ge m(s) K.
$$ 
with 
$$
m(s) = \min_{c'=1,\dots,C} \max_{c=1,\dots,C} \left\{ \alpha_c g_{c,c'} e^{- {s \over g_{0,c}}} \right\} > 0.
$$
where $m(s) > 0$ from assumption~\ref{ass:connect}.

For $i=1,\dots,K$, $X_i(s)$ is a sum of $K$ random variables bounded by $G$, so that Hoeffding's inequality yields:
$$
	\PP( |X_i(s) - \EE(X_i(s))| \ge G \sqrt{K  \ln K} ) \le {1 \over K^2}.
$$
Using a union bound over $i=1,\dots,K$:
\begin{align*}
	\PP( & \max_{i=1,\dots,K} |X_i(s) - \EE(X_i(s))| \ge G \sqrt{K \ln K} ) \\
	& \hspace{-0.5cm} \le \sum_{i=1}^K \PP( |X_i(s) - \EE(X_i(s))| \ge G \sqrt{K \ln K} ) \le {1 \over K} \toK 0.
\end{align*}
Therefore the event $${\cal A} = \{ \max_{i=1,\dots,K} |X_i(s) - \EE(X_i(s))| \le G \sqrt{K \ln K} \}$$ occurs with high probability when $K \to \infty$. Consider $K$ large enough so that $m(s) K -\sqrt{K \ln K} \ge 0$. If $\cal A$ occurs:
\begin{align*}
	\sum_{i=1}^K {| X_i(s) -  \EE(X_i(s))| \over X_i(s)\EE(X_i(s))} &\le  { K \sqrt{K \ln K} \over m(s) K (m(s)K -\sqrt{K \ln K})} \\
	&\toK 0.
\end{align*}

\underline{Second term} Compute the second moment of the second term:
\begin{align*}
	 \EE \left[ \Big( \sum_{i=1}^K { Z_i(s)  - e^{- {s \over \gamma_{0,i}}} \over \EE(X_i(s))} \Big)^2\right] &= \sum_{i=1}^K {e^{- {s \over \gamma_{0,i} }}(1 - e^{- {s \over \gamma_{0,i} }}) \over \EE(X_i(s))^2} \\ 
	 & \hspace{-1cm} \le \sum_{i=1}^K {1 \over \EE(X_i(s))^2} \le {1 \over m(s)^2 K} \toK 0,
\end{align*}
since $\EE( Z_i(s)) = e^{- {s \over \gamma_{0,i}} }$ and $Z_1(s),\dots,Z_K(s)$ are independent. So the second term goes to $0$ in probability, since $L^2$ convergence implies convergence in probability.

\underline{Putting it together} We have proven that:
\begin{align*}
		\sum_{i=1}^K { 1 - Z_i(s) \over X_i(s)}  &\PtoK   \sum_{i=1}^K {1 - e^{- {s \over \gamma_{0,i}}} \over \EE(X_i(s))} \\
		&= \sum_{c=1}^C {\alpha_c(1 - e^{-{s \over g_{0,c}}}) \over \sum_{c'=1}^C \alpha_{c'} g_{c',c} e^{- {s \over g_{0,c'}}} }.
	\end{align*}
Since convergence in probability implies convergence in distribution and $x \mapsto e^{-x}$ is continuous and bounded on $[0,\infty)$ we obtain the result:
$$
	 P_{+}(s) \toK \exp\left\{ - s \sum_{c=1}^C {\alpha_c(1 - e^{-{s \over g_{0,c}}}) \over \sum_{c'=1}^C \alpha_{c'} g_{c',c} e^{- {s \over g_{0,c'}}} }\right\}.
$$
\ep
\medskip
\begin{corollary}\label{cor:main_result_outage}
	When $K \to \infty$, the outage rate converges ${1 \over 2} \log_2(1 + s)$ where $s$ is the unique solution to: 
	\begin{align*}
		 s \sum_{c=1}^C {\alpha_c(1 - e^{-{s \over g_{0,c}}})  \over \sum_{c'=1}^C \alpha_{c'} g_{c',c} e^{- {s \over g_{0,c'}}} } = \ln\Big({1 \over 1 - \epsilon}\Big).
	\end{align*}
	and for $\epsilon \approx 0$ we have $s \approx 0$ and a Taylor development gives:
	$$
		s \approx \sqrt{ \ln\Big({1 \over 1-\epsilon}\Big)  \over  \sum_{c=1}^C {\alpha_c \over g_{0,c}}  \Big( \sum_{c'=1}^C \alpha_{c'} g_{c',c} \Big)^{-1}   }
	$$
\end{corollary}
\section{Numerical Experiments}\label{sec:numerical}

In this section we present numerical experiments and show that our theoretical analysis provides accurate predictions of the system's behavior. For each figure, three curves are presented: ''baseline'' is the performance of the baseline scheme, ''simulation'' is the exact performance of the proposed scheme obtained by simulation, and ''approx'' is the analytical approximation of the proposed scheme's performance. For the multicast rate the approximation is given by Theorem~\ref{th:main_res_multicast_approx} and for the outage rate, the approximation is given by Theorem~\ref{th:main_result_outage}. We consider three scenarios:

(a) A single class $C=1$, with $\alpha = (1)$ , $g_{0,1} = 46 $ dBm and $g_{1,1} = 23$ dBm.

(b) Two classes $C=2$, with $\alpha = (0.5,0.5)$, $g_{0,1} = 46$ dBm,$g_{0,2} = 0$, $g_{1,1} = g_{2,2} = 23$ dBm and $g_{1,2} = g_{2,1} = 13$ dBm.

(c) A cell of radius $250$ m, with $K$ users drawn uniformly at random in this area. The mean channel gains are taken as $\gamma_{i,j} = \rho_i - 128 - 36.4 \log_{10}(d_{i,j})$ dBm where $d_{i,j}$ is the distance between $i$ and $j$, expressed in kilometers, and $\rho_i= 46$ dBm if $i=0$ and $\rho_i= 23$ dBm otherwise. For this scenario, the results are averaged over $100$ random realizations.

\begin{figure}[h!]
    \centering
    \begin{subfigure}[b]{0.49\columnwidth}
        \includegraphics[width=\textwidth]{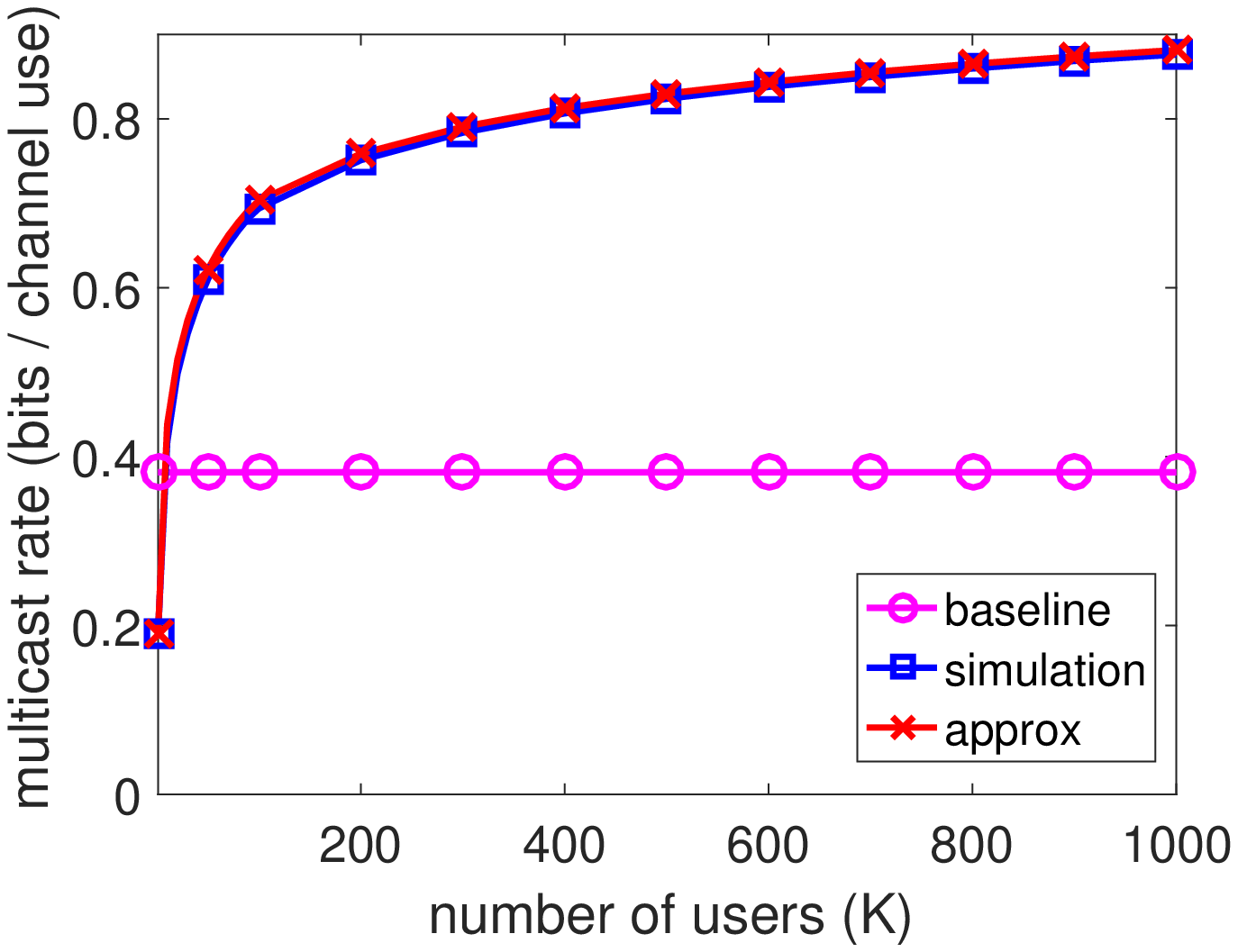}
        \caption{Scenario (a)}
        \label{fig:1a}
    \end{subfigure}
    \begin{subfigure}[b]{0.49\columnwidth}
        \includegraphics[width=\textwidth]{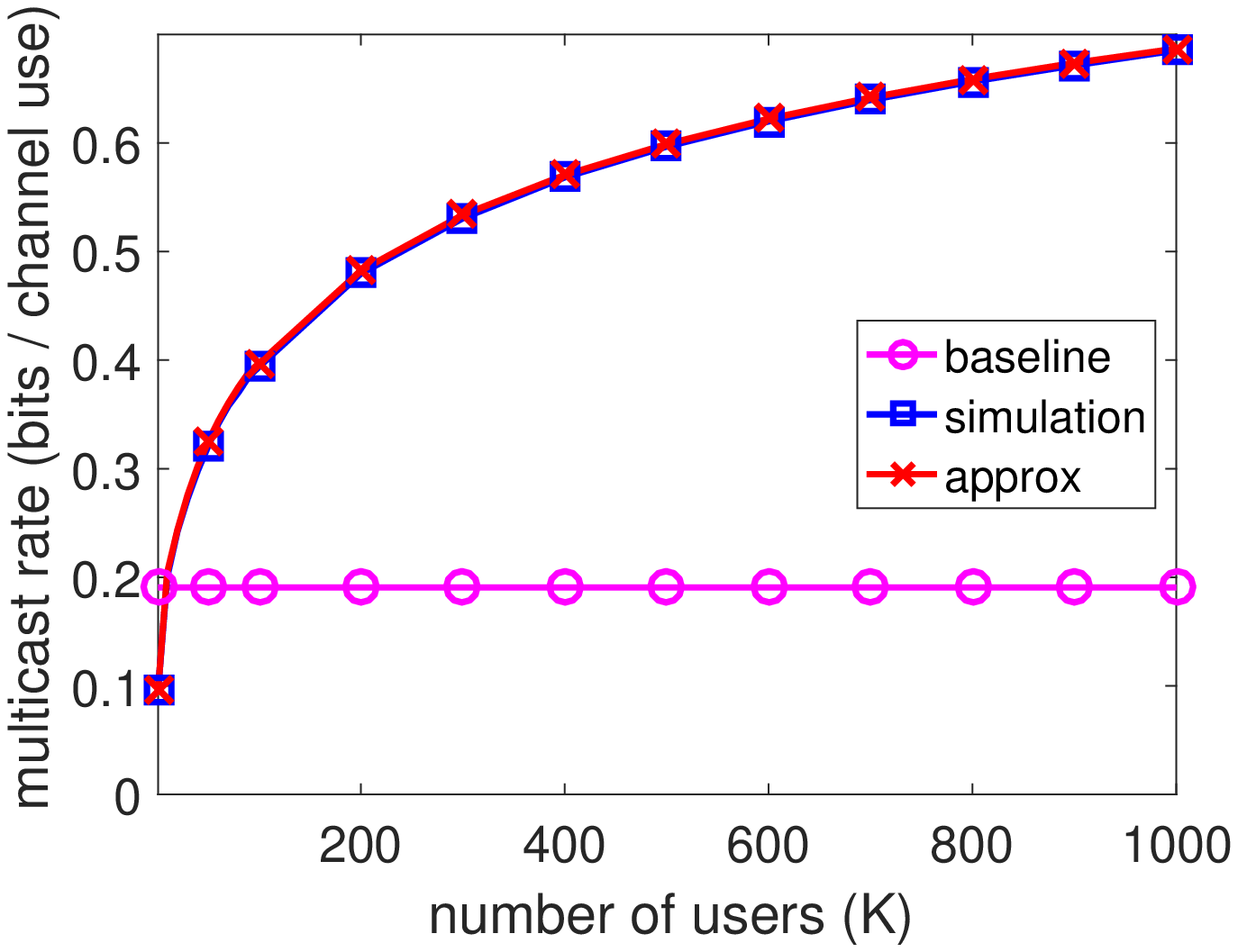}
        \caption{Scenario (b)}
        \label{fig:1b}
    \end{subfigure}
    
       \vspace{0.5cm}
     \begin{subfigure}[b]{0.49\columnwidth}
        \includegraphics[width=\textwidth]{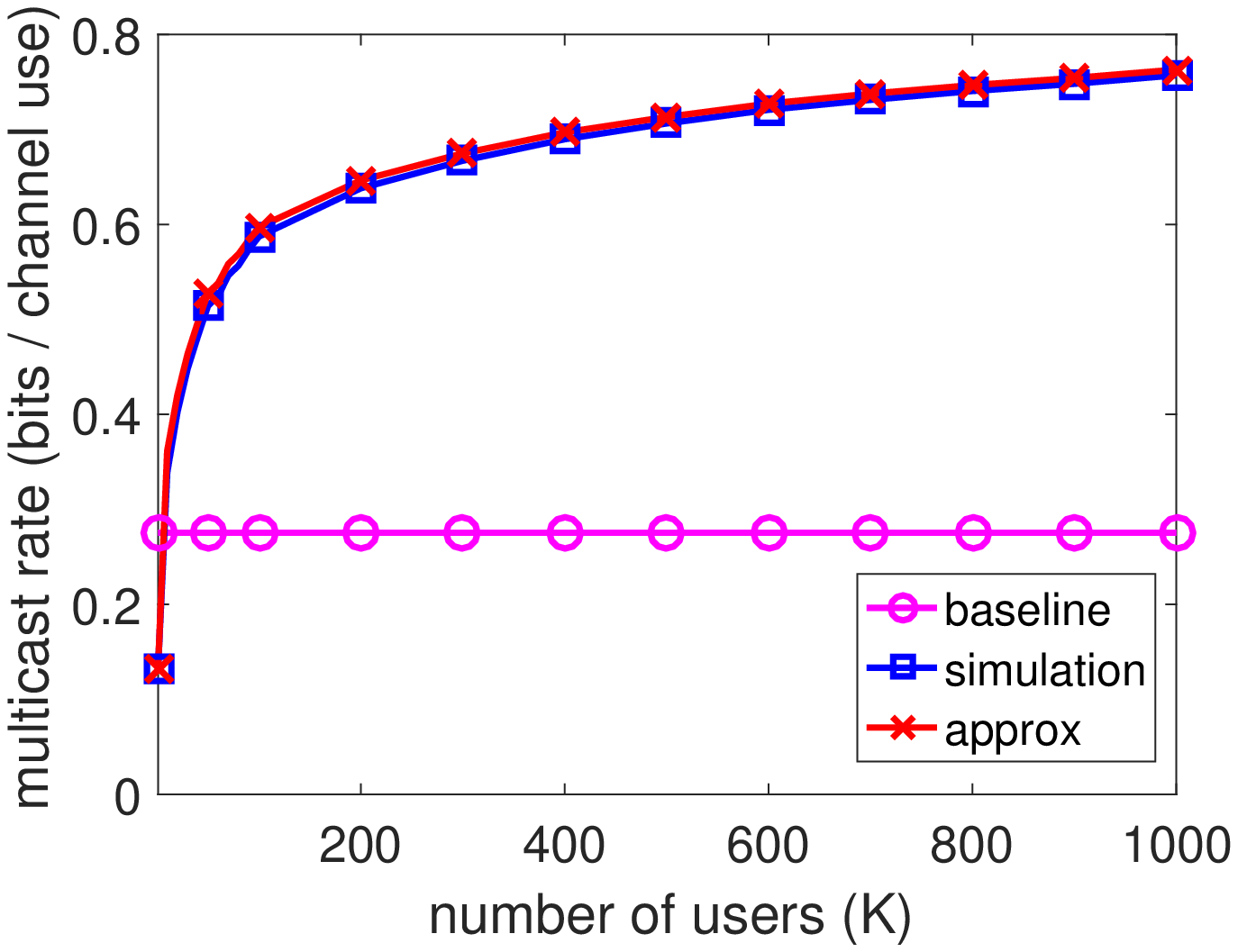}
        \caption{Scenario (c)}
        \label{fig:1c}
    \end{subfigure}   
    \caption{Multicast rate $R^m$ versus number of users $K$}\label{fig:1}
\end{figure}

\begin{figure}[h!]
    \centering
    \begin{subfigure}[b]{0.49\columnwidth}
        \includegraphics[width=\textwidth]{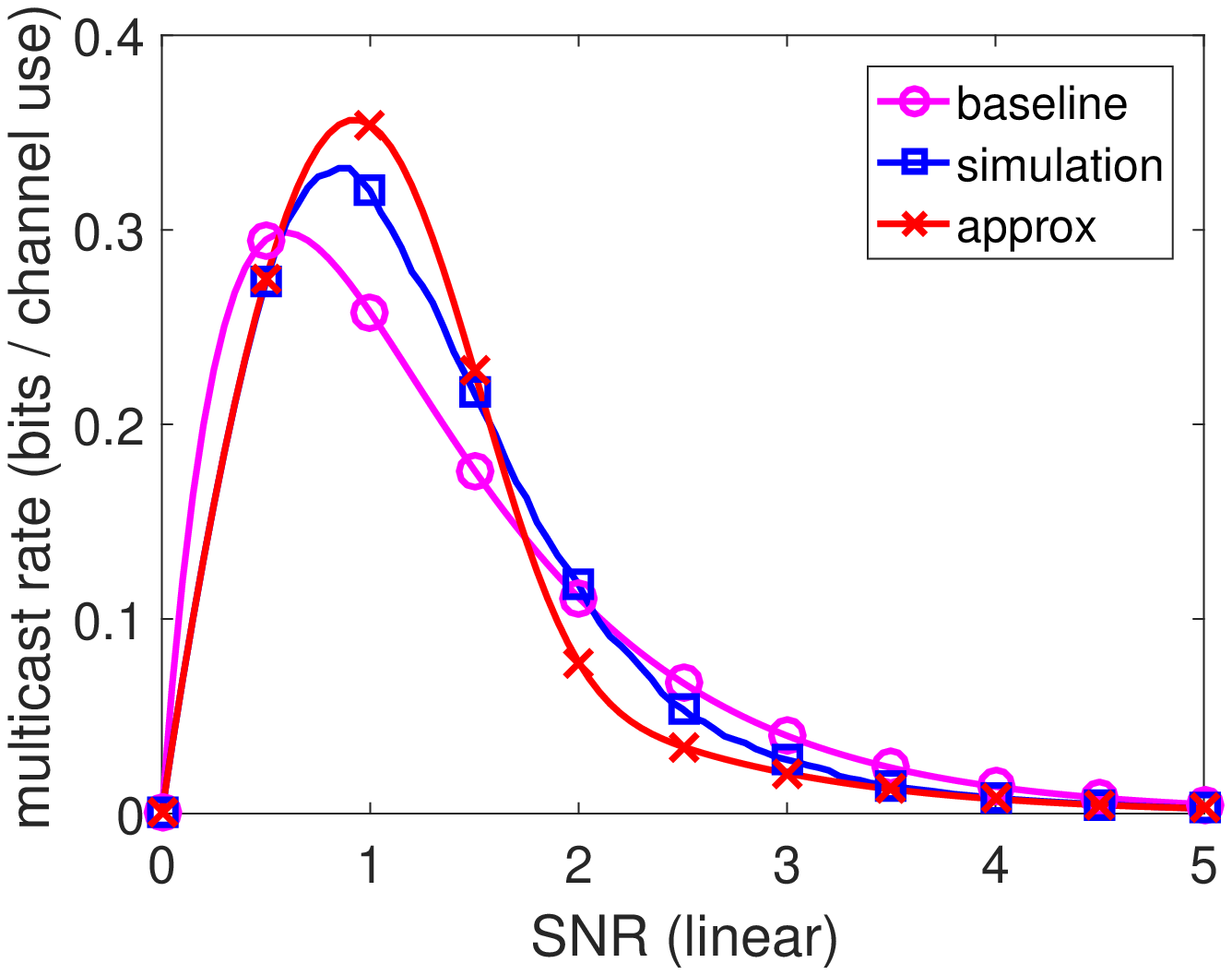}
        \caption{Scenario (c), $K=10$}
        \label{fig:4a}
    \end{subfigure}
    \begin{subfigure}[b]{0.49\columnwidth}
        \includegraphics[width=\textwidth]{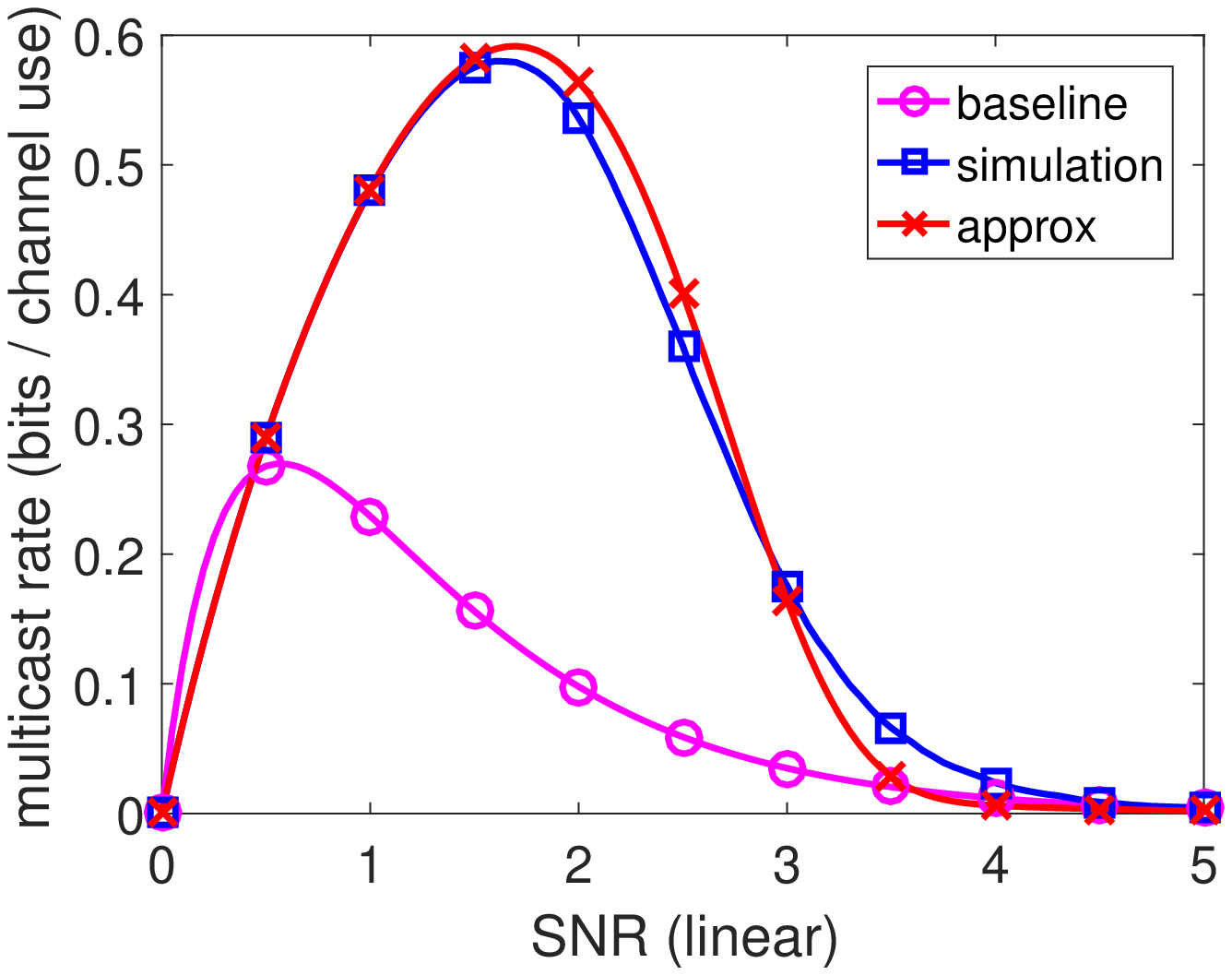}
        \caption{Scenario (c), $K=10^2$}
        \label{fig:4b}
    \end{subfigure}
    
     \vspace{0.5cm}
     \begin{subfigure}[b]{0.49\columnwidth}
        \includegraphics[width=\textwidth]{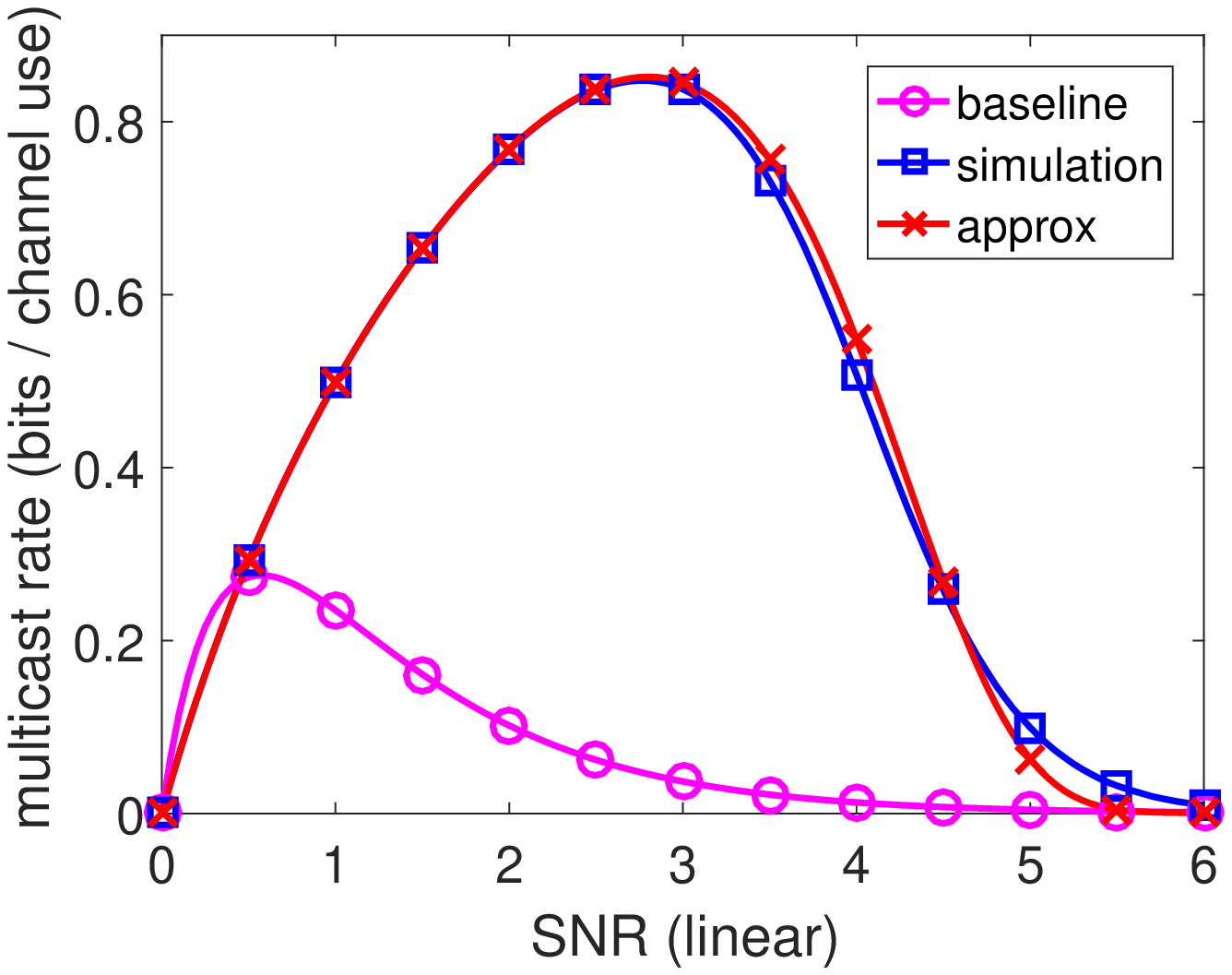}
        \caption{Scenario (c), $K=10^3$}
        \label{fig:4c}
    \end{subfigure}   
    \caption{Multicast, effective rate ${1 \over 2} \log_2(1+s) \bar{P}(s)$ vs SNR $s$}\label{fig:4}
\end{figure}

\begin{figure}[h!]
    \centering
    \begin{subfigure}[b]{0.49\columnwidth}
        \includegraphics[width=\textwidth]{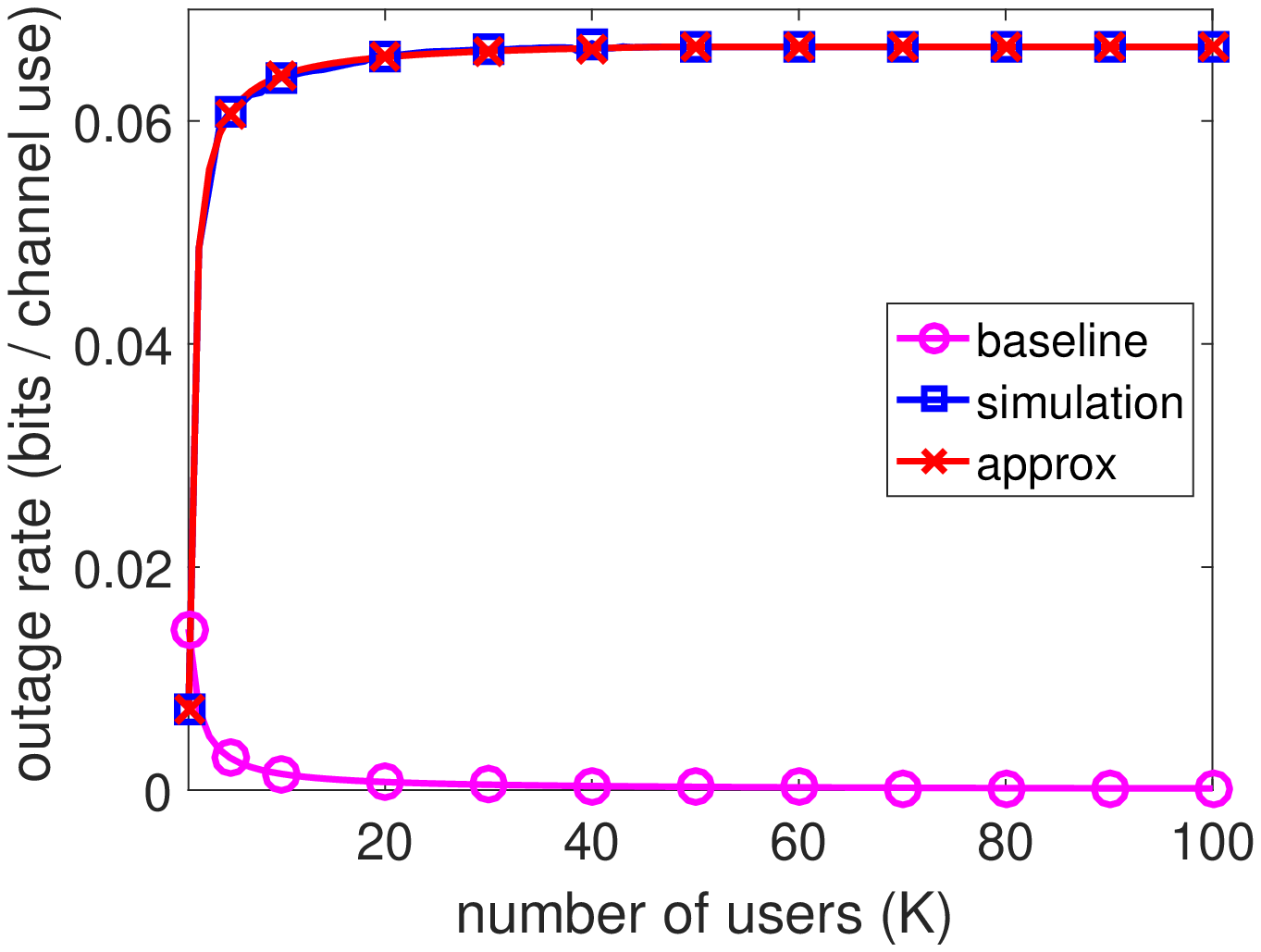}
        \caption{Scenario (a)}
        \label{fig:2a}
    \end{subfigure}
    \begin{subfigure}[b]{0.49\columnwidth}
        \includegraphics[width=\textwidth]{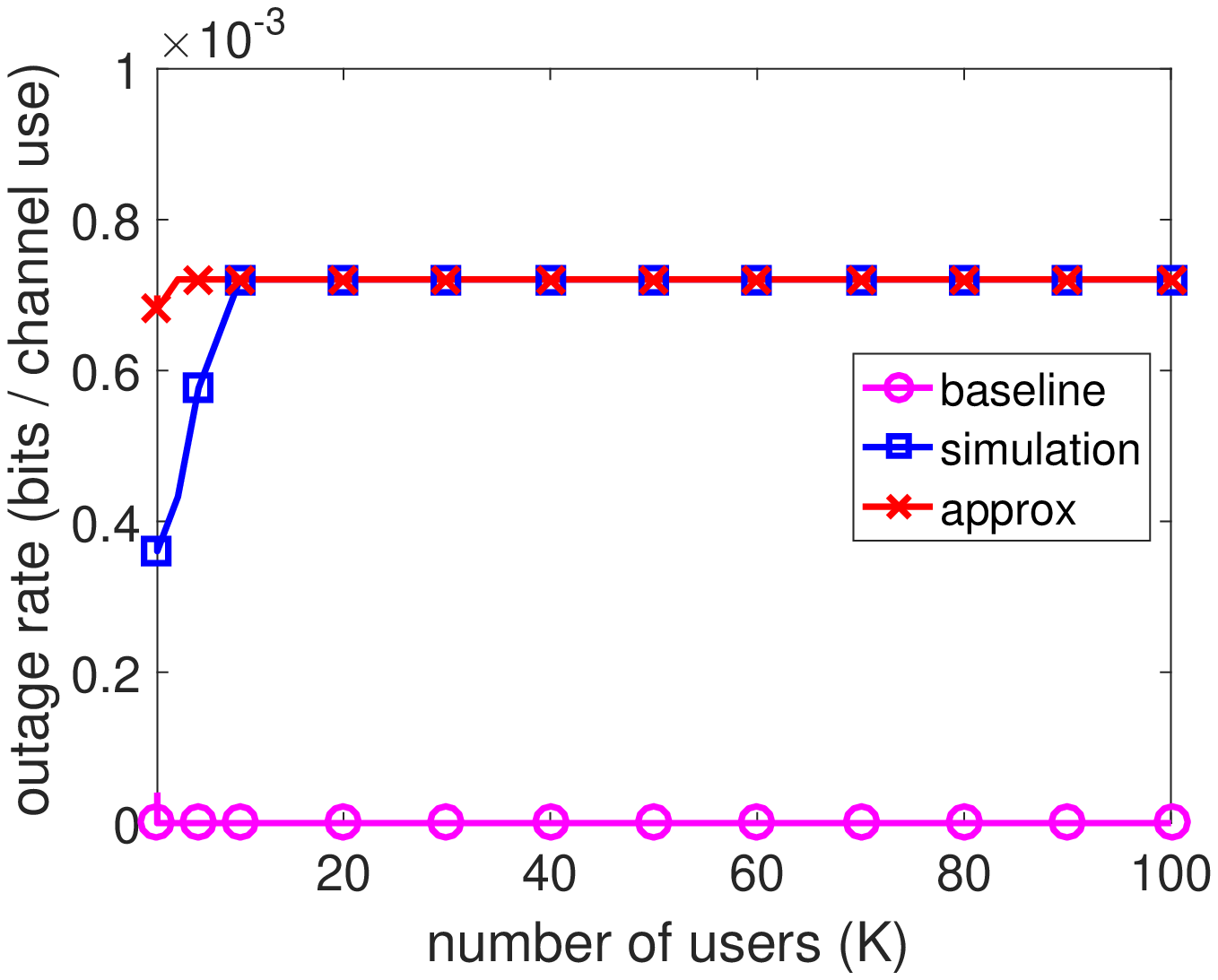}
        \caption{Scenario (b)}
        \label{fig:2b}
    \end{subfigure}
    
      \vspace{0.5cm}
     \begin{subfigure}[b]{0.49\columnwidth}
        \includegraphics[width=\textwidth]{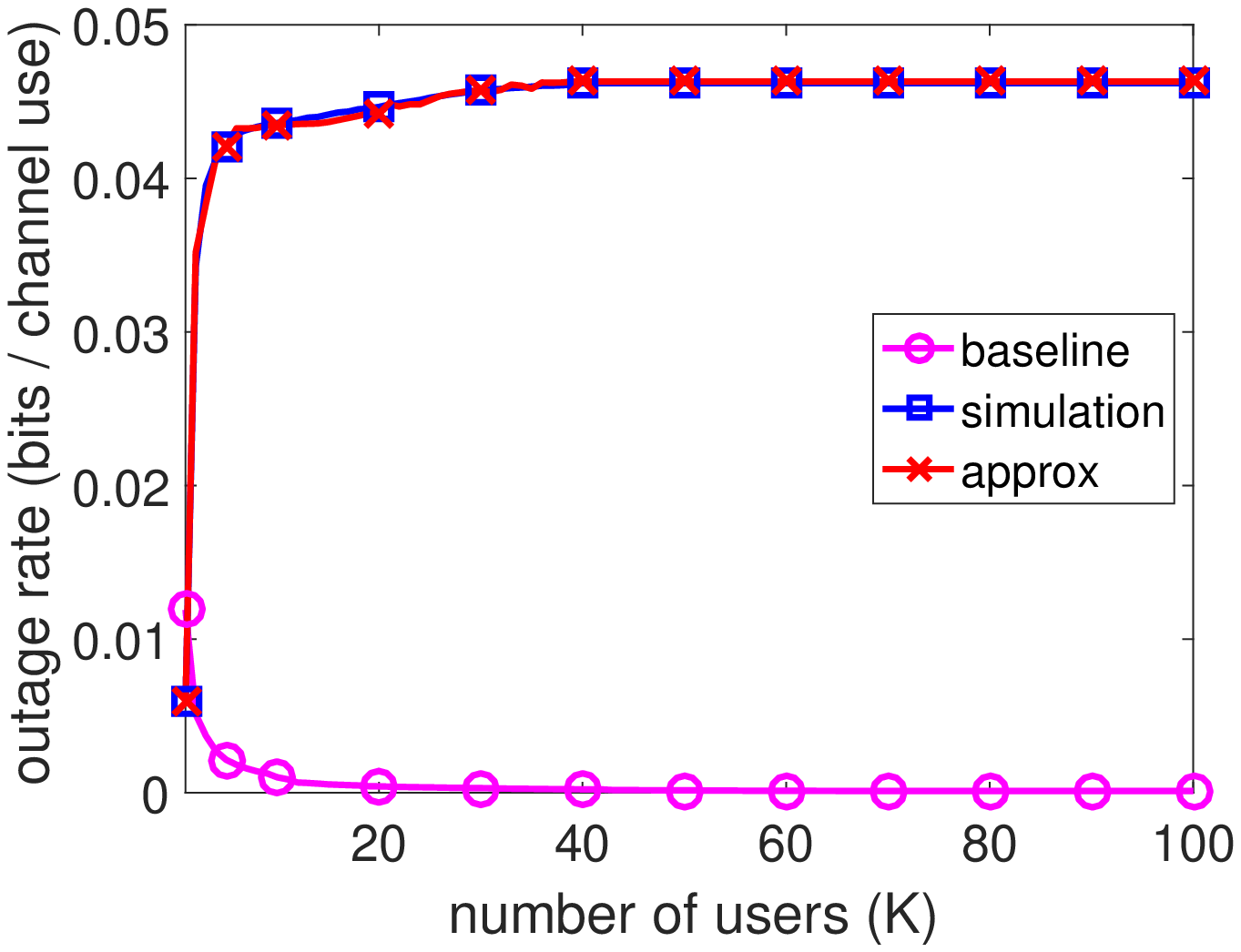}
        \caption{Scenario (c)}
        \label{fig:2c}
    \end{subfigure}   
    \caption{Outage rate $R^o$ versus number of users $K$}\label{fig:2}
\end{figure}

\begin{figure}[h!]
    \centering
    \begin{subfigure}[b]{0.49\columnwidth}
        \includegraphics[width=\textwidth]{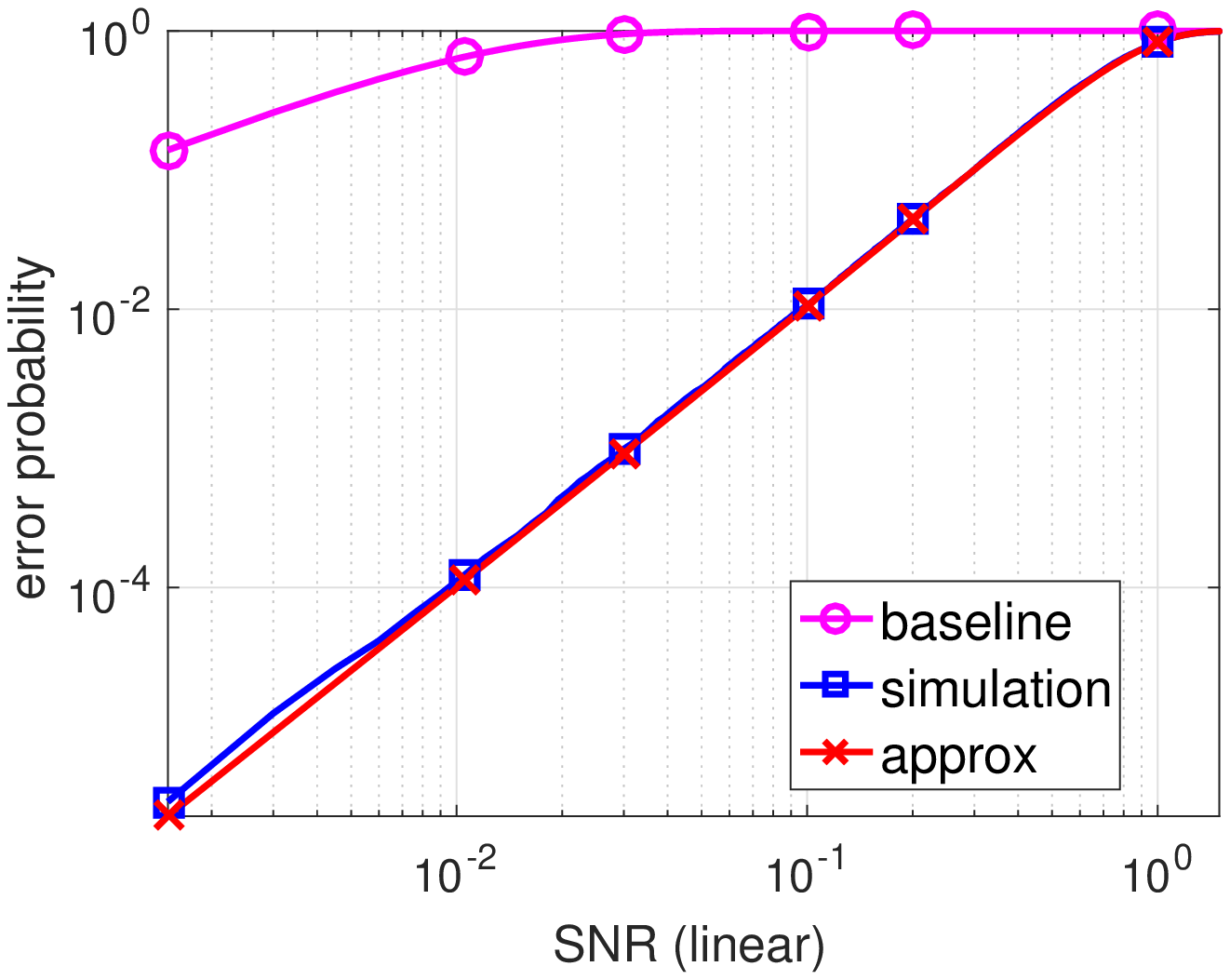}
        \caption{Scenario (a), $K = 10^2$}
        \label{fig:3a}
    \end{subfigure}
    \begin{subfigure}[b]{0.49\columnwidth}
        \includegraphics[width=\textwidth]{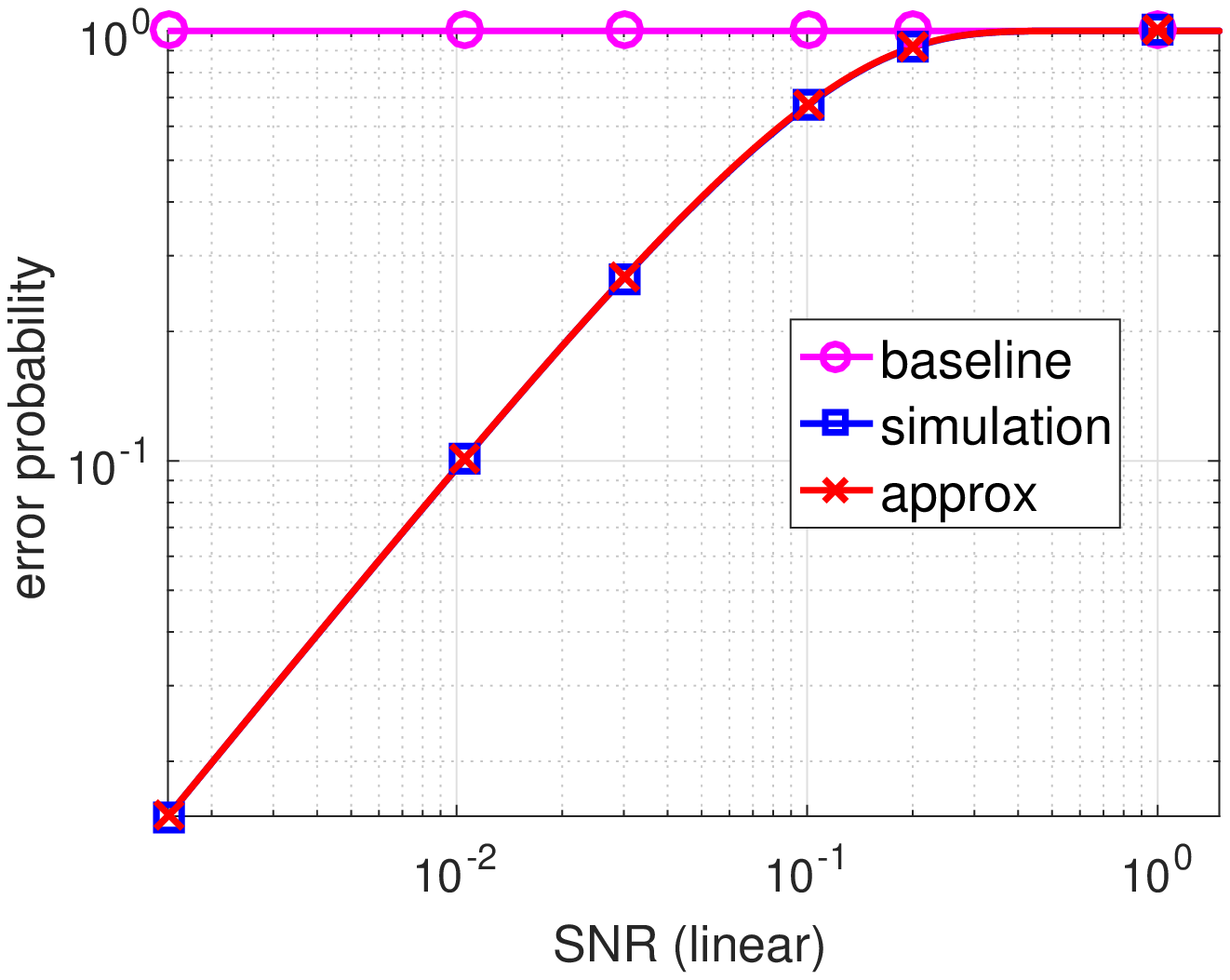}
        \caption{Scenario (b), $K = 10^2$}
        \label{fig:3b}
    \end{subfigure}
    
    \vspace{0.5cm}
     \begin{subfigure}[b]{0.49\columnwidth}
        \includegraphics[width=\textwidth]{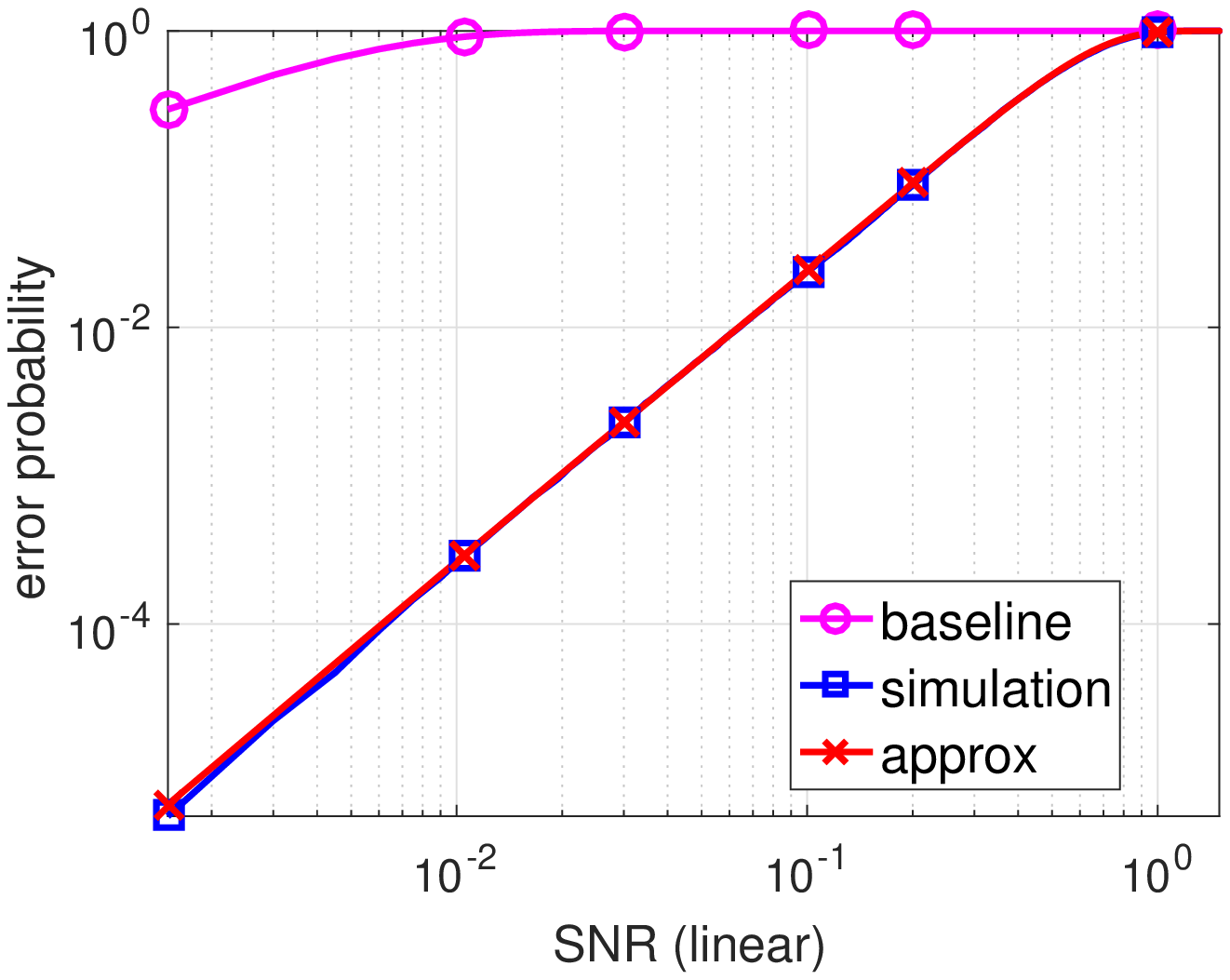}
        \caption{Scenario (c), $K = 10^2$}
        \label{fig:3c}
    \end{subfigure}   
    \caption{Outage rate, probability of error $1-P_{+}(s)$ vs SNR $s$}\label{fig:3}
\end{figure}

Scenario (a) is the homogeneous case where all users are close to the station and to each other, scenario (b) is a case where users of class $1$ are close to the station, and users of class $2$ are far away from the station, so that in order to receive data they require users of class $1$ to act as a relay. Finally scenario (c) represents the case where users arrive uniformly in a cell area, which seems like an acceptable model for a real system in high load (i.e. when there are many active users). For all scenarios $\epsilon = 10^{-2}$. For the multicast rate, figure~\ref{fig:1} presents the multicast rate $R^m$ as a function of the number of users $K$ (when $s$ is chosen optimally), while figure~\ref{fig:4} presents the multicast rate as a function of $s$, for various values of $K$. For outage rate, figure~\ref{fig:2} presents the outage rate as a function of the number of users $K$, while figure~\ref{fig:3} presents the outage probability $P_{+}(s)$ as a function of $s$ for system size $K = 100$.

These numerical experiments point out three facts:
\begin{itemize}
\item The proposed scheme is vastly superior to the baseline scheme as soon as there are more than a few users, as predicted by our asymptotic analysis.
\item The proposed approximations/asymptotic expressions derived above predict the performance of the proposed scheme very accurately, even for systems of modest size, say $K \ge 40$ for the multicast rate and $K \ge 15$ for the outage rate.
\item As a consequence, setting the parameter $s$ can be done in a simple and tractable manner to obtain good practical performance for systems of modest size.
\end{itemize}	
\section{Conclusion}\label{sec:conclusion}

We have studied a new scheme for broadcasting a message to users with receiver-to-receiver communication, without channel state information at the transmitter. We have shown that, in the large system limit, the performance of this scheme can be completely characterized. For finite systems, we also have provided tractable approximations for performance measures which are very accurate even for systems of modest sizes.  
\appendix

\section{Auxiliary results}
We recall below a few basic concentration inequalities. Proofs may be found, for instance, in~\cite{Lugosi}.
\begin{proposition}\label{prop:chebychev}[Chebychev's inequality]
	Consider $Y$ a random variable. Then for all $\delta > 0$ we have:
	\begin{align*}
		\PP(  |Y - \EE(Y)| \ge \delta) \le  {\var(Y) \over \delta^2}. 
	\end{align*}
\end{proposition}

\begin{proposition}\label{prop:hoeffding}[Hoeffding's inequality]
	Consider $Y_1,\dots,Y_K$ independent random variables in $[0,1]$. Then for all $\delta > 0$ we have:
	\begin{align*}
		\PP\left(  {1 \over K} \sum_{i=1}^K (Y_i - \EE(Y_i)) \ge \delta\right) \le e^{-2 K \delta^2}. 
	\end{align*}
\end{proposition}

\begin{proposition}\label{prop:chernoff}[Chernoff's inequality]
	Consider $Y_1,\dots,Y_K$ independent random variables in $[0,1]$. Define $y_K =  \sum_{i=1}^K \EE(Y_i)$. Then for all $\delta \in (0,1)$ we have:
	\begin{align*}
		\PP\left( \sum_{i=1}^K Y_i \ge (1+\delta) y_K \right) \le e^{-{y_K \delta^2 \over 3}}. 
	\end{align*}
	and
	\begin{align*}
		\PP\left( \sum_{i=1}^K Y_i \le (1-\delta) y_K \right) \le e^{-{y_K \delta^2 \over 3}}. 
	\end{align*}
\end{proposition}


\bibliographystyle{IEEEtran}
\bibliography{REF}

%
%
%
%
%
%
%

\end{document}